\documentclass[reviewer,10pt,authoryear,fleqn]{elsarticle}
\usepackage[margin=1in]{geometry}
\usepackage{amsmath}
\usepackage{amssymb}
\usepackage{graphicx}
\usepackage{hyperref}

\journal{New Astronomy}

\begin{document}

\begin{frontmatter}
\title{Effective sound speed in relativistic accretion discs around Schwarzschild black holes}	
\author[add1]{Md Arif Shaikh}
\ead{arifshaikh@hri.res.in}
\author[add1]{Susovan Maity}
\ead{susovanmaity@hri.res.in}
\author[add2]{Sankhasubhra Nag}
\ead{sankha@sncwgs.ac.in}
\author[add1,add3]{Tapas Kumar Das}
\ead{tapas@hri.res.in}

\address[add1]{Harish-Chandra Research Institute, HBNI, Chhatnag Road, Jhunsi, Allahabad 211 019, India}
\address[add2]{Sarojini Naidu College for Women, Kolkata 700 028, India}
\address[add3]{Physics and applied mathematics unit, Indian Statistical Institute, Kolkata 700 108, India}

\begin{abstract}
\noindent For low angular momentum axially symmetric accretion flow maintained in hydrostatic equilibrium along the vertical direction, the value of the Mach number at the critical points deviates from unity, resulting in the non-isomorphism of the critical and the sonic points. This introduces several undesirable complexities while analytically dealing with the stationary integral accretion solutions and the corresponding phase portraits. We propose that the introduction of an effective dynamical sound speed may resolve the issue in an elegant way. We linear perturb the full spacetime-dependent general relativistic Euler and the continuity equations governing the structure and the dynamics of accretion disc in vertical equilibrium around Schwarzschild black holes and identify the sonic metric embedded within the stationary background flow. Such metric describes the propagation of the linear acoustic perturbation inside the accretion flow. We construct the wave equation corresponding to that acoustic perturbation and find the speed of propagation of such perturbation. We finally show that the ordinary thermodynamic sound speed should be substituted by the speed of propagation of the linear acoustic wave which has been obtained through the dynamical perturbation. Such substitution will make the value of Mach number at the critical point to be equal to unity. Use of the aforementioned effective sound speed will lead to a modified stationary disc structure where the critical and the sonic points will be identical.
\end{abstract}

\end{frontmatter}

\section{Introduction}
\noindent Accretion flows onto astrophysical black holes are supposed to exhibit transonic properties in general \citep{Liang1980,Frank1985,Kato2008}. For low angular momentum, practically inviscid, axially symmetric accretion, sonic transition may take place at more than one locations on the equatorial plane of the disc and such multi-transonic flow may accommodate steady, standing shock transition \citep{Liang1980,Abramowicz1981,Muchotrzeb1982,Muchotrzeb1983,Fukue1983,Fukue1987,Lu1985,Lu1986,Muchotrzeb1986,Abramowicz1989,Chakrabarti1989,Abramowicz1990,CHAKRABARTI_physics_Reports,Kafatos1994,Yang1995,Pariev1996,Peitz1997,Caditz1998,Das2002,Das2003,Barai2004,Fukue2004,Abraham2006,Das2007,Okuda2004,Okuda2007,Das2012,Sukova2015JOP,Sukova2015MNRAS,Sukova2017}. Properties of the shocked multi-transonic accretion are usually studied for three different geometrical configurations of accreting matter, see, e.g, \citep{Chakrabarti2001,Abramowicz1990,Nag2012,Tarafdar2015,Tarafdar2018} for the details of such geometric configurations.\\

\noindent Among those three, one particular configuration, namely the accretion in hydrostatic equilibrium along the vertical direction, exhibits certain peculiar features. For such flow geometry, the Mach number at the critical points of the flow may not become unity \citep{Matsumoto1984viscous,Fukue1987,Das2002,Das2007,Das2012,Tarafdar2018} and hence the critical points may not be considered as sonic points. For accretion under the influence of various post-Newtonian pseudo-Schwarzschild or pseudo-Kerr black hole potentials, critical points for polytropic flow are formed at a location different from that of sonic points. For isothermal accretion under the influence of post-Newtonian black hole potentials, critical points and sonic points are, however, isomorphic. The amount of deviation of the value of the Mach number from unity, remains the same for both the saddle type sonic points for multi-transonic shocked polytropic flows under the influence of post-Newtonian potentials, and such deviations depends only on $\gamma$, where $\gamma$ is the ratio of the specific heats evaluated at constant pressure and at constant volume, respectively. For general relativistic accretion in the Schwarzschild or the Kerr metric, even for the isothermal flow the sonic point and the critical point can be located at two different radial coordinates on the equatorial plane as measured from center of the accretor. The amount of deviation of the value of the Mach number (evaluated at the critical point) from unity, may be found to be different for two different sonic points for multi-transonic flows.\\

\noindent Such non-isomorphism of critical points and sonic points, i.e, the fact that their locations may differ, may introduce various complexities while dealing with the multi transonic flow profile and related astrophysical phenomena. While plotting the stationary transonic integral solutions onto the Mach number versus radial distance phase portrait, phase orbits corresponding to the inwardly directed accretion and outward directed wind solutions intersect at the critical point. If the location of the critical point and its corresponding sonic point form at different locations, the subsonic and supersonic branches are found not to be identical with two branches of the phase orbits located at two sides of the critical points.\\

\noindent The critical points are obtained using the critical point analysis method -- a technique borrowed from the dynamical systems theory. For many of the accretion scenarios, it may be possible to locate the critical points analytically (see \citep{Agarwal2012} and references therein). Using certain eigenvalue techniques, one becomes able to gain, completely analytically, qualitative ideas about the phase portrait of the transonic flow structure close to the critical point \citep{Ray2003,Chaudhury2006,Goswami2007,Chaverra2016,Mandal2007}. If a sonic point is located at a distance different from that of the critical point, one needs to numerically integrate the flow equations, starting from the critical point, up to that particular point where the Mach number becomes unity. The elegance of the analytical eigenvalue-based methods is thus lost if a critical point and a sonic point are different. One needs to take recourse only to the complicated numerical techniques to have ideas about the subsonic and supersonic branches in the phase plot; see, e.g, for a very detail description of such numerical techniques.\\

\noindent Apart from the astrophysical point of view, accreting black hole systems have been studied from the perspective of emergent gravity phenomena \citep{Das2004,Dasgupta2005,Abraham2006,Das2007,Pu2012,Bilic2014,Tarafdar2015,Saha2016,Shaikh2017,Tarafdar2018,Shaikh2018a,Shaikh2018b} to understand how such systems can be perceived as an interesting example of classical analogue model naturally found in the universe. For such work also, the non-isomorphism between the critical and the sonic point may enhance the complexity involved with the solution scheme. The Mach number at the acoustic horizons should necessarily be unity, which requires the introduction of the numerical solution scheme to obtain the integral stationary flow solutions. Had it been the situation that the Mach number would be unity at the critical point an elegant analytical method could perhaps be employed to compute the value of acoustic surface gravity and related quantities, evaluated at sonic horizons.\\

\noindent The aforementioned discussions demand that it is imperative to introduce certain effective sound speed for which the effective Mach number evaluated at the critical points would be unity and the critical points and sonic points will be isomorphic. This will greatly reduce the complexity involved in employing numerical solution schemes for construction of the phase portrait and many other quantities relevant to the astrophysics of transonic black hole accretion and analogue gravity phenomena. The concept of effective sound speed has been discussed in the literature for accretion flows under the influence of post-Newtonian pseudo-Schwarzschild black hole potentials \citep{Matsumoto1984viscous,Fukue2004}. In the present work, we will provide a novel perturbative approach to introduce the concept of effective dynamical sound speeds embedded within the general relativistic, axially symmetric accretion flow maintained in the hydrostatic equilibrium along the vertical direction.\\

\noindent We consider three different expressions for disc thickness as proposed by \citet{Novikov-Thorne1973}, by \citet{Riffert-Herold1995} and by \citet{Abramowicz1996ap} to describe the accretion disc in hydrostatic equilibrium along the vertical direction in the Schwarzschild metric. For each of these three disc heights, we construct the time-independent Euler and the continuity equations. We solve such equations to find the corresponding first integrals of motion. For polytropic accretion, such first integrals are the total specific energy and the mass accretion rate $\Psi_0$. The polytropic accretion is parametrized by the specific energy $\xi_0^{\rm ad}$, the specific angular momentum $\lambda_0$, and the adiabatic index $\gamma$. A three-parameter set $[\xi_0^{\rm ad} ,\lambda_0 ,\gamma]$ where $\xi_0^{\rm ad} ,\lambda_0 ,\gamma$ are all constants, is taken to describe the flow and to solve the corresponding flow equations. For the isothermal accretion, two first integrals of motion are quasi-specific energy $\xi_0^{\rm iso}$(which is the integral solution of the time-independent part of the relativistic Euler equation) and the mass accretion rate $\Psi_0$. An isothermal flow is parametrized by $[T,\lambda_0 ]$, where $T$ and $\lambda_0$ are the conserved flow temperature and the constant specific angular momentum, respectively.\\

\noindent For all these three disc heights, we calculate for polytropic flow, the space gradient of the dynamical velocity and stationary sound speed, i.e., ${du_0}/{dr}$ and ${dc_{s0}}/{dr}$, respectively. From the expressions for ${du_0}/{dr}$ and ${dc_{s0}}/{dr}$, we evaluate the critical point conditions and compute the value of Mach number at the critical point. We show that the value of Mach number at the critical point is not unity and write down what would be the effective sound speed for which the Mach number at the critical point would have resumed the value unity.\\

\noindent We then linear perturb the full time-dependent Euler and continuity equation. Such perturbations lead to the formation of a space-time metric, which will be called the acoustic metric. The acoustic metric governs the dynamics of propagation of linear perturbation inside the background fluid (the fluid which composes the accretion disc). We then construct the corresponding wave equation for the propagation of such linear perturbation and calculate the speed of propagation of linear acoustic perturbation. We finally show that if we substitute the usual stationary sound speed $ c_s $ (as defined in equation (\ref{cs-ad})) by the suitable form of the speed of propagation of linear acoustic perturbation, then the Mach number at the critical points becomes unity. Hence we establish that certain representation of the `dynamical sound speed' (the speed of propagation of linear acoustic perturbation obtained through the dynamical stability analysis of the full spacetime-dependent fluid equations) should actually be considered as the effective speed of sound propagation along the equatorial plane of the black hole accretion disc. If one replaces the usual static sound speed by the aforementioned effective dynamical sound speed, the critical points always coincide with the sonic points and all the complexities originating from the non-isomorphism of the critical and the sonic points get resolved.\\

\noindent For isothermal flow, we perform the same operation for finding the Mach number at the critical points corresponding to the three different disc heights. We find that unlike the adiabatic accretion for which the accretion disc characterized by all three disc heights would produce a mismatch between the critical and the sonic points, for isothermal flow, accretion disc characterized by only one expression of disc height (proposed by \citet{Abramowicz1996ap}) produces the non-isomorphism between the critical and the sonic points. For accretion characterized by the other two expressions of disc heights (as proposed by \citet{Novikov-Thorne1973,Riffert-Herold1995}), the location of the critical and the sonic points are found to be the same.\\

\noindent In section \ref{Sec:Gov-eq}, we present the basic equations governing the general relativistic accretion flow and introduce relevant thermodynamic quantities. In section \ref{Sec:Crit-points}, we find out the conditions for critical points for the three different disc models of vertical equilibrium for adiabatic as well as the isothermal equation of state. In section \ref{Sec:Acoustic-metric}, we derive the acoustic spacetime metric by linear perturbing the accretion flow equations. Finally in section \ref{Sec:effective-speed}, using the acoustic spacetime metric we obtain the effective speed, $ c_{s0}^{\rm eff} $, of the propagation of the acoustic perturbations. This suggests that at the critical point one always have $ u_0^2 = {c_{s0}^{\rm eff}}^2 $.

We shall set $G=c=M_{\rm BH}=1$ where $G$ is the universal gravitational constant, $c$ is the velocity of light and $M_{\rm BH}$ is the mass of the black hole. The radial distance will be scaled by $G M_{\rm BH}/c^2$ and any velocity will be scaled by $c$. We shall use the negative-time-positive-space metric convention.
\section{Governing equations}\label{Sec:Gov-eq}
\noindent We consider an inviscid axially symmetric irrotational accretion flow accreting onto a Schwarzschild black hole. The background spacetime metric could be written in the following form
\begin{equation}\label{key}
ds^2 = -g_{tt}dt^2+g_{rr}dr^2+g_{\theta\theta}d\theta^2+g_{\phi\phi}d\phi^2,
\end{equation}
where the metric elements are given by
\begin{equation}\label{key}
g_{tt} = g_{rr}^{-1} = (1-{2}/{r}),\quad g_{\theta\theta} = {g_{\phi\phi}}/{\sin^2\theta} = r^2.
\end{equation}
The energy momentum tensor for a perfect fluid is given by
\begin{equation}\label{energy-mom-tensor}
T^{\mu\nu} = (p+\varepsilon)v^\mu v^\nu + p g^{\mu\nu},
\end{equation}
where $ p $ is the pressure and $ \rho $ is the rest-mass energy density of the fluid. $ \varepsilon $ is the total energy density of the fluid which is the sum of the rest-mass energy density and the thermal energy density, i.e., $ \varepsilon = \rho + \varepsilon_{\rm thermal} $. $ v^\mu $ is the four-velocity with the normalization condition $ v^\mu v_\mu = -1 $. The equation of state for adiabatic flow is given by $ p = k\rho^\gamma $ where $ k $ is a constant. Whereas for isothermal case $ p\propto \rho $. The sound speed for adiabatic flow (isoentropic flow) is given by
\begin{equation}\label{cs-ad}
c_{s}= \left.\frac{dp}{d\varepsilon}\right|_{\rm entropy = constant} = \frac{\rho}{h}\frac{dh}{d\rho},
\end{equation}
where $ h $ is the enthalpy given by 
\begin{equation}\label{enthalpy}
h = \frac{p+\varepsilon}{\rho}.
\end{equation}
On the other hand the sound speed for isothermal flow can be defined as \citep{Yuan1996}
\begin{equation}\label{cs-iso}
c_{s}^2 = \frac{1}{h}\frac{dp}{d\rho},
\end{equation}
where $ h = {\rm constant} $ for isothermal case.\\

\noindent The mass conservation equation and the energy-momentum conservation equations are given by, respectively,
\begin{equation}\label{continuity}
\nabla_\mu (\rho v^\mu) = 0,
\end{equation}
and 
\begin{equation}\label{energy-mom-con}
\nabla_\mu T^{\mu\nu} = 0.
\end{equation}
Using the expression for the sound speed the energy momentum conservation equation can be written in the following form
\begin{equation}\label{Euler}
v^\mu \nabla_\mu v^\nu +\frac{c_{s}^2}{\rho}(v^\mu v^\nu + g^{\mu\nu})\partial_\mu \rho = 0,
\end{equation}
where $ c_s $ for adiabatic case and isothermal case are given by equation (\ref{cs-ad}) and equation (\ref{cs-iso}), respectively.

\section{Accretion disc models and critical points}\label{Sec:Crit-points}
\noindent To find the critical points of the accretion flow, we have to find the expression of the gradient of the advective velocity $ u_0 $, i.e., the expression of $ du_0/dr $ for stationary accretion flow. In order to that we need two constant integrals of the stationary flow. The first one comes from the continuity equation and the second one comes from the momentum conservation equation. It is convenient to do a vertical averaging of the flow equations by integrating over $ \theta $ and the resultant equation is described by the flow variables defined on the equatorial plane ($ \theta = \pi/2 $). In addition one also integrates over $ \phi $ which gives a factor of $ 2\pi $ due to the axial symmetry of the flow. We do such vertical averaging as prescribed in \citep{Gammie1998} to the continuity equation given by Eq. (\ref{continuity}). Thus in case of stationary ($ t $-independent) and axially symmetric ($ \phi $-independent) flow with averaged $ v^\theta \sim 0 $, the continuity equation can be written as
\begin{equation}\label{continuity-avg-st}
\frac{\partial}{\partial r}(4\pi H_\theta\sqrt{-\tilde{g}}\rho_0 v_0^r) = 0,
\end{equation}
where the factor $ H_\theta $ arises due to the vertical averaging and is the local angular scale of flow. Thus one can relate the actual local flow thickness $ H(r) $ to the angular scale of the flow $ H_\theta $ as $ H_\theta = H(r)/r $, where $ r $ is the radial distance along the equatorial plane from the center of the disc. $ \tilde{g} $ is the value of the determinant of the metric $ g_{\mu\nu} $ on the equatorial plane, $\tilde{g} = {\rm det}(g_{\mu\nu})|_{\theta=\pi/2} = -r^4$. The equation (\ref{continuity-avg-st}) gives the mass accretion rate $ \Psi_0 $ as
\begin{equation}\label{Sationary-mass-acc-rate}
\Psi_0 = 4\pi \sqrt{-g}H_\theta  \rho_0 v_0^r = 4\pi H(r) r \rho_0 v_0^r.
\end{equation}
The $ t,r $ component of the four velocity, $v^t, v^r $, can be expressed in terms of $ u_0 $ and $ \lambda_0 = -v_{\phi 0}/v_{t0} $ as, respectively,\citep{Gammie1998}
\begin{equation}\label{transformation}
\begin{aligned}
& v^t_0 = \sqrt{\frac{g_{\phi\phi}}{g_{tt}(g_{\phi\phi}-\lambda_0^2 g_{tt})}}\frac{1}{\sqrt{1-u_0^2}},\\ 
& v^r_0 = \frac{u_0}{\sqrt{g_{rr}(1-u_0^2)}} = \frac{\sqrt{\Delta}u_0}{r\sqrt{1-u_0^2}}.
\end{aligned}
\end{equation}
Using $ g_{rr}=r^2/\Delta $, with $ \Delta = r(r-2) $. $ \lambda_0 $ is the specific angular momentum of the fluid and is a constant for stationary flow. Thus $ \Psi_0 $ can be written as
\begin{equation}\label{Psi}
\Psi_0 = 4\pi H(r)\Delta^{1/2}\rho_0 \frac{u_0}{\sqrt{1-u_0^2}}.
\end{equation}
For adiabatic flow, we define a new quantity $ \dot{\Xi }$ from $ \Psi_0 $ by multiplying it with $ (\gamma k)^{\frac{1}{\gamma-1}} $. $ \dot{\Xi} $ is a measure of entropy accretion rate and typically called as the entropy accretion rate. Expressing $ \rho_0 $ in terms of $ \gamma, k$ and $c_{s0} $ finally gives
\begin{equation}\label{Xi}
\dot{\Xi} = \left(\frac{c_{s0}^2}{1-nc_{s0}^2}\right)^n4\pi H(r)\Delta^{1/2} \frac{u_0}{\sqrt{1-u_0^2}} = {\rm constant},
\end{equation}
where we have used $ n=1/(\gamma-1) $. The second conserved quantity can be obtained from the time-component of the relativistic Euler equation (\ref{Euler})
which for stationary adiabatic case gives 
\begin{equation}
\xi^{\rm ad}_0 = - h_0v_{t0} = {\rm constant},
\end{equation}
and for stationary isothermal case gives
\begin{equation}
\xi^{\rm iso}_0 = -\rho_0^{c_{s0}^2}v_{t0} = {\rm constant},
\end{equation}
where $ c_{s0} $ is a constant for isothermal flow.
$ v_{t0} $ can be further expressed in terms of $ u_0 $ as

\begin{equation}
v_{t0} = -\sqrt{\frac{\Delta}{B(1-u_0^2)}},
\end{equation}
where $ B = g_{\phi\phi}-\lambda_0^2 g_{tt} $. Thus
\begin{equation}\label{xi}
\xi^{\rm ad}_0 = \frac{1}{1-nc_{s0}^2}\sqrt{\frac{\Delta}{B(1-u_0^2)}},
\end{equation}
and 
\begin{equation}\label{xiiso}
\xi^{\rm iso}_0 = \rho_0^{c_{s0}^2}\sqrt{\frac{\Delta}{B(1-u_0^2)}}.
\end{equation}
For adiabatic flow, the expression for $ du_0/dr $ can be derived by using the expression of the two quantities, $ \dot{\Xi} $ and $ \xi^{\rm ad}_0 $ given by equation (\ref{Xi}) and (\ref{xi}), respectively. Taking logarithmic derivative of both sides of equation (\ref{xi}) gives the gradient of sound speed as
\begin{equation}\label{dcdr}
\left.\frac{dc_{s0}}{dr}\right|^{\rm ad} = -\frac{1-nc_{s0}^2}{2nc_{s0}}\left[\frac{u_0}{1-u_0^2}\frac{du_0}{dr}+\frac{1}{2}\left(\frac{\Delta'}{\Delta}-\frac{B'}{B}\right)\right].
\end{equation}
For isothermal flow, we make use of equation (\ref{Psi}) and (\ref{xiiso}). Taking logarithmic derivative of the equation (\ref{xiiso}) we can find $ (d\rho/dr)/\rho_0 $ as
\begin{equation}\label{rho'}
\left.\frac{\frac{d\rho}{dr}}{\rho_0}\right|^{\rm iso} = -\frac{1}{c_{s0}^2}\left[\frac{u_0}{1-u_0^2}\frac{du_0}{dr}+\frac{1}{2}\left(\frac{\Delta'}{\Delta}-\frac{B'}{B}\right)\right].
\end{equation}
Below we discuss different models of vertical structure of accretion disc and the corresponding critical point conditions for stationary accretion flow in such model of accretion disc.

\subsection{Models of accretion disc under vertical equilibrium}
\label{sec:disc-models}
\noindent In the beginning of the current section we mentioned that for accretion disc flow, in order for the governing equation to be written in terms of the variables evaluated at the equatorial plane, the equations are vertically averaged which introduces the disc height $H(r)$ or equivalently the local angular scale of the flow $H_\theta$ in the resulting equations. Thus in order to solve for the accretion flow profile, we need to have an expression for the local thickness of the accretion disc. In our present work, we are concerned with accretion disc which is under hydrostatic equilibrium in the vertical direction. In Newtonian accretion flow, for accretion disc under vertical equilibrium, the disc height calculation is a rather straightforward work of balancing the pressure gradient in the vertical direction with the component of the Newtonian gravitational force in the vertical direction.\\

\noindent In case of a general relativistic accretion disc around a black hole, one needs to incorporate the general relativistic effects on the balancing of pressure gradient and gravitational force. Historically there have been three such general relativistic models of disc height which incorporated the general relativistic effects. The first of such prescriptions of disc height was given by \citet{Novikov-Thorne1973}. In deriving the expression for the disc height, \citet{Novikov-Thorne1973} replaced the Newtonian formula for acceleration by the vertical acceleration which is calculated from the Riemann tensor $R^{3}_{030}$ given in \citep{Bardeen1972} and transformed to the local tetrad. A relatively improved expression was given by \citet{Riffert-Herold1995} who derived the gravity-pressure balance equation itself by imposing two particular orthonormality condition on the vertical component of the Euler equation. However, both the disc models of \citet{Novikov-Thorne1973} and that of \citet{Riffert-Herold1995} do not apply below $r=3$ (in the units we are working with) where the disc height becomes zero. Thus the disc height expressions are not valid up to the horizon $r =2$ (for a Schwarzschild black hole). \citet{Abramowicz1996ap} provided an expression for the disc height which is regular up to the horizon. \citet{Abramowicz1996ap} derived the equation directly from the relativistic Euler equation and no additional simplifying assumptions were made.\\

\noindent In the following, we work with the above mentioned three prescriptions of disc heights for general relativistic accretion disc under hydrostatic equilibrium in the vertical direction around Schwarzschild black holes.

\subsection{ Novikov-Thorne (NT)}
\noindent The expression for the disc height as derived by \citet{Novikov-Thorne1973} for accretion disc around Schwarzschild black hole could be given by
\begin{equation}\label{H_nt}
    H_{\rm NT}(r) = \sqrt{\frac{p_0}{\rho_0}}r^{3/2}\sqrt{\frac{r-3}{r-2}}.
\end{equation}
\subsubsection{Adiabatic case}
\noindent For adiabatic equation of state, $ p_0/\rho_0 $ can be written as 
\begin{equation}
    \frac{p_0}{\rho_0} = \left(\frac{n}{n+1}\right)\left(\frac{c_{s0}^2}{1-nc_{s0}^2}\right).
\end{equation}
Thus we can write $ H(r) $ as
\begin{equation}
    H_{\rm NT}(r) = \left(\frac{n}{n+1}\right)^{1/2}\left(\frac{c_{s0}^2}{1-nc_{s0}^2}\right)^{1/2}f_{\rm NT}(r),
\end{equation}
where $ f_{\rm NT}(r) = r^{3/2}\sqrt{(r-3)/(r-2)} $. Using this expression of $ H(r) $, $ \dot{\Xi} $ for this model can be written as
\begin{equation}
    \dot{\Xi}_{\rm NT} = \sqrt{\frac{n}{n+1}}\left(\frac{c_{s0}^2}{1-nc_{s0}^2}\right)^\frac{2n+1}{2}4\pi\Delta^{1/2} \frac{u_0}{\sqrt{1-u_0^2}}f_{\rm NT}(r).
\end{equation}
Taking logarithmic derivative of both sides of the above equation and substituting $ dc_{s0}/dr $ using Eq. (\ref{dcdr}) gives
\begin{equation}
    \left.\frac{du_0}{dr}\right|^{\rm ad}_{\rm NT} = \frac{u_0(1-u_0^2)\left[\frac{2n}{2n+1}c_{s0}^2(\frac{\Delta'}{2\Delta}+\frac{f_{\rm NT}'}{f_{\rm NT}})+\frac{1}{2}(\frac{B'}{B}-\frac{\Delta'}{\Delta})\right]}{u_0^2-\frac{c_{s0^2}}{1+\frac{1}{2n}}}=\frac{N^{\rm ad}_{\rm NT}}{D^{\rm ad}_{\rm NT}}.
\end{equation}
The critical points are obtained from the condition $ D^{\rm ad}_{\rm NT} = 0 $ which gives $ u_0^2|_c =c_{s0}^2/(1+(1/2n))|_c  $ or
\begin{equation}
u_0^2|_c = \frac{c_{s0}^2|_c}{1+\beta},\quad {\rm where}\quad \beta = \frac{\gamma-1}{2}.
\end{equation}

\subsubsection{Isothermal case}\label{Sec:iso-nt}
\noindent For isothermal equation of state, $ p = k_0\rho$ ($ k_0 $ is a constant), the disc height is given by 
\begin{equation}\label{key}
H^{\rm iso}_{\rm NT} = \sqrt{k_0}r^{3/2}\sqrt{\frac{r-3}{r-2}} = \sqrt{k_0}f_{\rm NT}(r).
\end{equation}
Therefore, the mass accretion rate is given by
\begin{equation}\label{key}
\Psi^{\rm iso}_{\rm NT} = 4\pi \sqrt{k_0}\Delta^{1/2}\rho_0 \frac{u_0}{\sqrt{1-u_0^2}}f_{\rm NT}(r).
\end{equation}
Taking  logarithmic derivative of the above equation with respect to $ r $ and substituting $ (d\rho_0/dr)/\rho_0 $ using equation (\ref{rho'}) gives

\begin{equation}\label{key}
\left.\frac{du_0}{dr}\right|_{\rm NT}^{\rm iso} = \frac{u_0(1-u_0^2)\left[c_{s0}^2\left(\frac{f'_{\rm NT}}{f_{\rm NT}}+\frac{\Delta'}{2\Delta}\right)+\frac{1}{2}\left(\frac{B'}{B}-\frac{\Delta'}{\Delta}\right)\right]}{u_0^2-c_{s0}^2} = \frac{N^{\rm iso}_{\rm NT}}{D^{\rm iso}_{\rm NT}}.
\end{equation}
Thus critical points are given by the condition $ D^{\rm iso}_{\rm NT} = 0 $, which gives 
\begin{equation}\label{key}
u_0^2|_c = c_{s0}^2|_c.
\end{equation}
\subsection{ Riffert-Herold (RH)}
\noindent \citet{Riffert-Herold1995} (RH) improved the result obtained by NT. The modified expression for the disc height is given by 
\begin{equation}\label{H_rh}
    H_{\rm RH}(r) = 2\sqrt{\frac{p}{\rho}} r^{3/2}\sqrt{\frac{r-3}{r}}.
\end{equation}    
\subsubsection{Adiabatic case}
\begin{equation}
H_{\rm RH}(r) = \left(\frac{n}{n+1}\right)^{1/2}\left(\frac{c_{s0}^2}{1-nc_{s0}^2}\right)^{1/2}f_{\rm RH}(r),
\end{equation}
where $ f_{\rm RH}=2r\sqrt{r-3} $. $H_{\rm RH}(r)$ has the same form as that of $ H_{\rm NT} $. Therefore, the expression for $ du_0/dr $ can be derived similarly which gives
\begin{equation}
    \left.\frac{du_0}{dr}\right|^{\rm ad}_{\rm RH} = \frac{u_0(1-u_0^2)\left[\frac{2n}{2n+1}c_{s0}^2(\frac{\Delta'}{2\Delta}+\frac{f_{\rm RH}'}{f_{\rm RH}})+\frac{1}{2}(\frac{B'}{B}-\frac{\Delta'}{\Delta})\right]}{u_0^2-\frac{c_{s0^2}}{1+\frac{1}{2n}}} = \frac{N^{\rm ad}_{\rm RH}}{D^{\rm ad}_{\rm RH}}.
\end{equation}
Setting $ D^{\rm ad}_{\rm RH} = 0 $ gives the critical point condition as
\begin{equation}
u_0^2|_c = \frac{c_{s0}^2|_c}{1+\beta},\quad {\rm where}\quad \beta = \frac{\gamma-1}{2}. 
\end{equation}

\subsubsection{Isothermal case}
\noindent Following the same procedure as given in section \ref{Sec:iso-nt}, $du_0/dr  $ for isothermal equation of state would be given by 

\begin{equation}\label{key}
\left.\frac{du_0}{dr}\right|_{\rm RH}^{\rm iso} = \frac{u_0(1-u_0^2)\left[c_{s0}^2\left(\frac{f'_{\rm RH}}{f_{\rm RH}}+\frac{\Delta'}{2\Delta}\right)+\frac{1}{2}\left(\frac{B'}{B}-\frac{\Delta'}{\Delta}\right)\right]}{u_0^2-c_{s0}^2} = \frac{N^{\rm iso}_{\rm RH}}{D^{\rm iso}_{\rm RH}},
\end{equation}
which gives the critical point condition as $ u_0^2 = c_{s0}^2 $.

\subsection{ Abramowicz-Lanza-Percival (ALP)}
\noindent The expression for the disc height as given by \citet{Abramowicz1996ap} can be written as
\begin{equation}\label{H_alp}
H(r) = \sqrt{2}r^2\sqrt{\frac{p}{\rho}}\frac{1}{|v_{\phi 0}|} = \sqrt{\frac{2}{\alpha}}\sqrt{\frac{p}{\rho}}\frac{r^2}{\lambda_0}\sqrt{1-u_0^2},
\end{equation} 
where we have used $ v_{\phi 0}= -\lambda_0 v_{t0} $. 

\subsubsection{Adibatic case}
\noindent Using the above expression for the disc height the entropy accretion rate can be written as
\begin{equation}\label{key}
\dot{\Xi}_{\rm ALP} = \sqrt{\frac{n}{n+1}}\left(\frac{c_{s0}^2}{1-nc_{s0}^2}\right)^{\frac{2n+1}{2}}\frac{r^2}{\lambda_0}4\pi \sqrt{2B}u_0.
\end{equation}
Taking logarithmic derivative of the above equation with respect to $ r $ and substituting $ dc_{s0}/dr $ using equation (\ref{dcdr}) gives the expression for the gradient of advective velocity as
\begin{equation}\label{dudr-ALP}
\left.\frac{du_0}{dr}\right|^{\rm ad}_{\rm ALP} = \frac{c_{s0}^2 u_0 (1-u_0^2)\left[\frac{B'}{2B}+\frac{2}{r}-\frac{2n+1}{4nc_{s0}^2}\left(\frac{\Delta'}{\Delta}-\frac{B'}{B}\right)\right]}{\left(\frac{2n+1}{2n}+c_{s0}^2\right)\left(u_0^2-\frac{c_{s0}^2}{1+\frac{1}{2n}+c_{s0}^2}\right)} = \frac{N^{\rm ad}_{\rm ALP}}{D^{\rm ad}_{\rm ALP}}.
\end{equation}
Setting $ D^{\rm ad}_{\rm ALP} = 0 $ gives the critical point condition as
\begin{equation}\label{key}
u_0^2|_c = \left.\frac{c_{s0}^2}{1+\beta}\right|_c,\quad \beta = \frac{\gamma-1}{2}+c_{s0}^2.
\end{equation}
\subsubsection{Isothermal case}
\noindent The disc height for isothermal case, with the equation of state $ p=k_0\rho $ , would be given by
\begin{equation}\label{key}
H(r) = \sqrt{\frac{2}{\alpha}} \sqrt{k_0}\frac{r^2}{\lambda_0}\sqrt{1-u_0^2}.
\end{equation}
For isothermal case, the mass accretion rate can be written as
\begin{equation}\label{key}
\left.\Psi\right|^{\rm iso}_{\rm ALP} = \rho_0\sqrt{k}\frac{r^2}{\lambda_0}4\pi \sqrt{2B}u_0.
\end{equation}
Taking  logarithmic derivative of the above equation with respect to $ r $ and substituting $ (d\rho_0/dr)/\rho_0 $ using equation (\ref{rho'}) gives
\begin{equation}\label{key}
\left.\frac{du_0}{dr}\right|^{\rm iso}_{\rm ALP} = \frac{c_{s0}^2u_0(1-u_0^2)\left[\frac{B'}{2B}+\frac{2}{r}-\frac{1}{2c_{s0}^2}\left(\frac{\Delta'}{\Delta}-\frac{B'}{B}\right)\right]}{(1+c_{s0}^2)(1-\frac{c_{s0}^2}{1+c_{s0}^2})} = \frac{N^{\rm iso}_{\rm ALP}}{D^{\rm iso}_{\rm ALP}}.
\end{equation}
Setting $ D^{\rm iso}_{\rm ALP} $ give the critical point condition as
\begin{equation}\label{key}
u_0^2 = \left.\frac{c_{s0}^2}{1+\beta}\right|_c,\quad \beta = c_{s0}^2.
\end{equation}
\noindent Thus we summarize the results obtained for different vertical equilibrium disc models and equations of states as follows. The critical points for any disc model and equation of state are obtained from the condition
\begin{equation}\label{key}
u_0^2|_c = \left.\frac{c_{s0}^2}{1+\beta}\right|_c,
\end{equation} where $ \beta $ depends on the disc model and the equation state. We give the values of $ \beta $ in table \ref{beta}.
\begin{table}[]
    \centering
    \begin{tabular}{p{6cm}p{3cm}p{3cm}}\hline
        Accretion disc models & $ \beta $ for isothermal equation of state & $ \beta $ for adiabaic equation of state \\ \hline
        Novikov \& Thorne (NT) & 0 & $ \frac{\gamma-1}{2} $\\ \hline
        Riffert \& Herold (RH) & 0 & $ \frac{\gamma-1}{2} $ \\ \hline
        Abramowicz Lanza \& Percival (ALP) & $ c_{s0}^2 $ & $ \frac{\gamma-1}{2}+c_{s0}^2 $ \\ \hline
    \end{tabular}
    \caption{Values of $ \beta $ for different disc structure models. Critical point condition is given by $u^2_0|_c =\frac{c_{s0}^2|_c}{{1+\beta}} $.}\label{beta}
\end{table}

\section{Acoustic spacetime metric}\label{Sec:Acoustic-metric}
\noindent In this section, we derive the acoustic spacetime metric by linear perturbing the equations governing the accretion flow. Following standard linear perturbation analysis, we write the time-dependent accretion variables, for example, the velocity components and density, as small time-dependent fluctuations about their stationary values. Therefore,
\begin{equation}\label{perturbations}
\begin{aligned}
& v^t(r,t) = v^t_0(r)+{v^t_1}(r,t),\\
& v^r(r,t) = v^r_0(r)+{v^r_1}(r,t),\\
& \rho(r,t) = \rho_0(r)+\rho_1(r,t),
\end{aligned}
\end{equation}
where the quantities with subscript `1' are the small time-dependent perturbations about the stationary quantity denoted by subscript `0'. We define a new variable $ \Psi = 4\pi \sqrt{-g}  \rho(r,t) v^r(r,t) H_\theta$ which is equal to the stationary mass accretion rate for the stationary accretion flow and hence
\begin{equation}\label{psi-perturbation}
\Psi(r,t) = \Psi_0 + \Psi_1(r,t),
\end{equation}
where $ \Psi_0 $ is the stationary mass accretion rate defined in equation (\ref{Sationary-mass-acc-rate}). The geometric factor $ 4\pi $ is just a constant and therefore, we can redefine the mass accretion rate $ \Psi $ as simply $ \Psi = \sqrt{-g} \rho(r,t)v^r(r,t) H_\theta$ without any loss of generality. Using the equations (\ref{perturbations}) we get 
\begin{equation}\label{Psi1}
\Psi_1 = \sqrt{-g}[\rho_1 v_0^r H_{\theta0}+\rho_0 v^r_1H_{\theta 0}+\rho_0 v_0^rH_{\theta 1}].
\end{equation}
It could be noticed that the perturbation $ \Psi_1 $ contains a term which is the perturbation of $ H_\theta $. We remember that $ H_\theta $ is the local angular scale of the flow and is related to the local flow thickness $ H(r) $ as $ H_{\theta} = H(r)/r$. The expressions for the disc thickness for vertical equilibrium model of Novikov-Thorne, Riffert-Herold and Abramowicz-Lanza-Percival are given by equation (\ref{H_nt}), (\ref{H_rh}) and (\ref{H_alp}), respectively. These expression contains $ p/\rho $ and further analysis needs an equation of state. Below we perform the analysis for adiabatic equation of state and isothermal equation of state.

\subsection{Acoustic metric for adiabatic flow}\label{Sec:Adiabatic}
\noindent For adiabatic flow, pressure is given by $ p = k\rho^\gamma $. The enthalpy given by equation (\ref{enthalpy}) can be thus written as 
\begin{equation}
h=1+\frac{\gamma}{\gamma-1}\frac{p}{\rho},
\end{equation}
and the perturbation $ h_1 $ can be written as
\begin{equation}\label{h1}
h_1 = \frac{h_0 c_{s0}^2}{\rho_0}\rho_1.
\end{equation}
We assume the accretion flow to be irrotational. Irrotationality condition provides the following equation for adiabatic flow \citep{Bilic1999}
\begin{equation}\label{key}
\partial_\mu(hv_\nu)-\partial_\nu(hv_\mu) = 0.
\end{equation}
The above equation along with the spherical symmetry of the flow (which implies $ \partial_\phi =0 $) provide the conserved quantity $ hv_\phi = {\rm  constant} $. Thus, using equation (\ref{h1}) one obtains
\begin{equation}\label{vphi1}
v_{\phi 1} = -\frac{v_{\phi 0}c_{s0}^2}{\rho_0}\rho_1.
\end{equation}
Linear perturbing the equation given by normalization condition, i.e., $ v_\mu v^\mu = -1 $ and using $ v^\phi_1 = (1/g_{\phi\phi})v_{\phi1} $ gives the perturbation of $ v^t $ in terms of $ v^r_1 $ and $ \rho_1 $ as 
\begin{equation}\label{vt1}
v_1^t=\alpha_1 v_1^r+\alpha_2 \rho_1, \quad \alpha_1 = \frac{g_{rr}v_0^r}{g_{tt}v_0^t}\quad{\rm and}\quad \alpha_2 = -\frac{g_{\phi\phi} (v_{\phi}^{0})^2 c_{s0}^2}{g_{tt}v_0^t \rho_0}.
\end{equation}
We express $ H_{\theta 1} $ in terms of perturbations of other quantities. For Novikov-Thorne and Riffert-Herold, we get
\begin{equation}
\frac{H_{\theta1}}{H_{\theta 0}} = \left(\frac{\gamma -1}{2} \right)\frac{\rho_1}{\rho_0},
\end{equation}
and for Abramowicz-Lanza-Percival, we have 
\begin{equation}
\frac{H_{\theta1}}{H_{\theta 0}} = \left( c_{s0}^2 +\frac{\gamma -1}{2} \right)\frac{\rho_1}{\rho_0}.
\end{equation}
Thus the expressions for different vertical equilibrium disc models can be given by an single equation as
\begin{equation}\label{h_theta1}
\frac{H_{\theta1}}{H_{\theta 0}} = \beta \frac{\rho_1}{\rho_0},
\end{equation}
where $ \beta $ for different disc models for adiabatic equation of state are given in Table \ref{beta}. Using this expression of $ H_{\theta1} $, we derive the acoustic metric for the three different disc models in a combined way.\\

\noindent The continuity equation for vertically averaged accretion flow takes the form 
\begin{equation}
\partial_t(\sqrt{-g}\rho v^t H_\theta)+\partial_r(\sqrt{-g}\rho v^r H_\theta) = 0.
\end{equation}
Using equation (\ref{perturbations}) and (\ref{psi-perturbation}) in the above equation and further using equation (\ref{vt1}) and (\ref{h_theta1}) provides
\begin{equation}\label{delrpsi}
\frac{\partial_r \Psi_1}{\Psi_0} = -\left[  \left\{ \frac{\alpha_2}{v_0^r}  + ( 1+ \beta ) \frac{v_0^t}{\rho_0 v_0^r} \right\} \partial_t \rho_1 + \frac{\alpha_1}{v_0^r} \partial_t v_1^r \right].
\end{equation}
Differentiating equation (\ref{Psi1}) with respect to $ t $ and using equation (\ref{h_theta1}) gives 
\begin{equation}\label{deltpsi}
\frac{\partial_t \Psi_1}{\Psi_0}= ( 1+\beta ) \frac{ \partial_t \rho_1}{\rho_0} + \frac{1}{v_0^r} \partial_t v_1^r.
\end{equation}
Equation (\ref{delrpsi}) and (\ref{deltpsi}) could be used to express $ \partial_t v^r_1 $ and $ \partial_t\rho_1 $ entirely in terms of derivatives of $ \Psi_1 $. This provides
\begin{equation}\label{deltv1}
\frac{\partial_t v_1^r}{v_0^r}=\frac{1}{\Lambda} \left[ \left\{ g_{tt} (v_0^t)^2 (1+ \beta ) - g_{\phi \phi}(v_0^\phi)^2 c_{s0}^2 \right\}\frac{\partial_t \Psi_1}{\Psi_0} +(1+\beta )g_{tt} v_0^r v_0^t \frac{\partial_r \Psi_1}{\Psi_0} \right],
\end{equation}
and 
\begin{equation}\label{deltrho1}
\frac{\partial_t \rho_1}{\rho_0}= -\frac{1}{\Lambda} \left[g_{tt} (v_0^r)^2\frac{\partial_t \Psi_1}{\Psi_0} + g_{tt} v_0^r v_0^t \frac{\partial_r \Psi_1}{\Psi_0} \right],
\end{equation}
where $ \Lambda $ is given by
\begin{equation}\label{Lambda}
 \Lambda = (1+\beta)+(1+\beta-c_{s0}^2)g_{\phi\phi}(v^\phi_0)^2.
\end{equation}
The temporal component of the Euler equation (\ref{Euler}) for axially symmetric flow can be written as
\begin{equation}
v^t\partial_t v^t +\frac{c_s^2}{\rho}\frac{ \{ g_{rr}(v_r)^2+g_{\phi\phi}(v_\phi)^2 \} }{g_{tt}}\partial_t\rho+v^r v^t \partial_r \left\{ \ln(hv_t)\right\}=0.
\end{equation}
Differentiating the above equation with respect to $ t $ and using the perturbation equations (\ref{perturbations}), (\ref{h1}) and (\ref{vt1}) provides
\begin{equation}
\partial_t \left(\frac{\alpha_1}{v_0^r}\partial_t v_1^r \right) +\partial_t \left( \frac{\alpha_1 c_s^2}{\rho_0}\partial_t \rho_1\right) + \partial_r \left(\frac{\alpha_1}{v_0^t}\partial_t v_1^r \right) + \partial_r \left\{ \left( \frac{ \alpha_2 }{v_0^t} +\frac{c_s^2}{\rho_0}\right)\partial_t \rho_1 \right\}=0.
\end{equation}
Finally substituting $ \partial_t v^r_1 $ and $ \partial_t \rho_1 $ in the above equation using equation (\ref{deltv1}) and (\ref{deltrho1}), respectively, gives the following equation
\begin{equation}\label{wave-eq}
\partial_\mu(F^{\mu\nu}\partial_\nu \Psi_1) = 0,
\end{equation}
where $ \mu,\nu $ run from $ 0 $ to $ 1 $. $ 0 $ stands for $ t $ and $ 1 $ stands for $ r $. The matrix $ F^{\mu\nu} $ is symmetric and is given by
\begin{equation}\label{key}
F^{\mu\nu} = \frac{g_{rr}v_0 c_{s0}^2}{v^t_0 \Lambda}\begin{bmatrix}
-g^{tt}+(1-\frac{1+\beta}{c_{s}^2}) (v^t_0)^2 & v^r_0 v^t_0(1-\frac{1+\beta}{c_{s}^2})  \\
v^r_0 v^t_0(1-\frac{1+\beta}{c_{s}^2})  & g^{rr}+(1-\frac{1+\beta}{c_{s}^2}) 
\end{bmatrix}.
\end{equation}
The equation (\ref{wave-eq}) describes the propagation of the perturbation $ \Psi_1 $. Equation (\ref{wave-eq}) mimics the wave equation of a massless scalar field $ \varphi $ in curved spacetime (with metric $ g^{\mu\nu} $) given by
\begin{equation}\label{scalarfield}
    \partial_\mu(\sqrt{-g}g^{\mu\nu}\partial_\nu \varphi)=0,
\end{equation}
where $ g $ is the determinant of the metric $ g_{\mu\nu} $. Comparing equation (\ref{wave-eq}) and (\ref{scalarfield}) one obtains the acoustic spacetime $ G^{\mu\nu} $ metric as
\begin{equation}\label{key}
\sqrt{-G}G^{\mu\nu} = F^{\mu\nu},
\end{equation}
where $ G $ is the determinant of $ G_{\mu\nu} $. Thus the acoustic metric $ G_{\mu\nu} $ would be given by
\begin{equation}\label{Gmunu}
G_{\mu\nu} = k(r) \begin{bmatrix}
-g^{rr}-(1-\frac{1+\beta}{c_{s0}^2})(v^r_0)^2 & v^r_0 v^t_0(1-\frac{1+\beta}{c_{s0}^2})  \\
v^r_0 v^t_0(1-\frac{1+\beta}{c_{s0}^2})  & g^{tt}-(1-\frac{1+\beta}{c_{s0}^2}) (v^t_0)^2
\end{bmatrix},
\end{equation}
where $ k(r) $ is some conformal factor arising due to the process of inverting $ G^{\mu\nu} $ in order to obtain $ G_{\mu\nu} $. For our current purpose we do not need to find the exact expression for $ k(r) $. In the following section we will be using equation (\ref{Gmunu}) to solve for the null acoustic geodesic.

\subsection{Acoustic metric for isothermal flow}
\noindent The procedure to derive the acoustic metric for isothermal flow is exactly the same as laid out in the previous section \ref{Sec:Adiabatic}. The differences comes from the difference in the equation of state. For isothermal equation of state, $ p/\rho = {\rm constant} $. The sound speed is defined by equation (\ref{cs-iso}). The irrotationality condition for isothermal flow is given by \citep{Shaikh2017}
\begin{equation}\label{key}
\partial_\mu(\rho^{c_{s}^2}v_\nu) - \partial_\nu(\rho^{c_{s}^2}v_\mu) = 0,
\end{equation}
where the sound speed $ c_s $ is a constant for isothermal flow. Using the above equation and the axial symmetry of the flow provides
\begin{equation}\label{key}
\rho^{c_{s}^2}v_\phi = {\rm constant}.
\end{equation}
Linear perturbation of the above equation leads to the same equation as equation (\ref{vphi1}) with $ c_s $ now a constant. The perturbation of $ H_\theta $ gives
\begin{equation}\label{key}
\frac{H_{\theta1}}{H_{\theta0}} = \beta \frac{\rho_1}{\rho_0},
\end{equation}
where $ \beta $ for isothermal flow for different model is given is Table \ref{beta}. The detailed derivation of acoustic metric for isothermal flow for vertical equilibrium model of ALP could be found in \citep{Shaikh2017}. The other models follows the same. This leads to the acoustic metric which has the same form as given in equation (\ref{Gmunu}) with the only difference is that the sound speed $ c_s $ is a constant for the isothermal case.

\section{Effective speed of acoustic perturbation}\label{Sec:effective-speed}
\noindent The acoustic metric for general relativistic axially symmetric disc in Schwarzschild spacetime was derived in the previous section for adiabatic and isothermal flow which is given by equation (\ref{Gmunu}). From the acoustic metric given by equation (\ref{Gmunu}), one can find out the location of the acoustic horizon. In analogy to the black hole event horizon in general relativity, the acoustic horizon can be defined as a null surface which acts like a one-way membrane for the acoustic perturbation. In other words, the acoustic perturbations inside the acoustic horizon cannot escape to the outside. For transonic flow, the transonic surface where bulk velocity and speed of acoustic perturbations becomes equal should act like such horizon. Because once the matter flow becomes supersonic, any acoustic perturbations will be dragged along the medium and hence the perturbation cannot escape to the subsonic region. If a surface $ r = {\rm constant} $ is the horizon, then the condition that the surface is null with respect to the metric $ G_{\mu\nu} $ provides \footnote{Such identification of the event horizon is only possible in specific spacetime geometries where the spacetime metric is stationary, asymptotically flat and have spherical topology. For details see \citep{carroll2004}}
\begin{equation}\label{key}
G_{\mu\nu} n^{\mu}n^{\nu} = 0,
\end{equation}
where $ n_{\mu}  = \delta_\mu^r$ is the normal to the surface $ r = {\rm constant} $ \citep{Moncrief1980}. Thus $ G^{rr} = 0  $ gives the location of the sonic horizon. Therefore, using the equation (\ref{transformation}), the location of the sonic or acoustic horizon is given by $ u_0^2 = c_{s0}^2/(1+\beta) $. However, as argued earlier, the acoustic horizon is basically the transonic surface which in turn implies that the effective speed of the acoustic perturbation is $ c_{s0}^{\rm eff} = c_{s0}/\sqrt{1+\beta} $. Now, in section \ref{Sec:Crit-points}, we showed that the critical point conditions for the different models and equations of state can be written in a single form as $ u_0^2 = c_{s0}^2/(1+\beta) $. Therefore, the critical point condition becomes $ u_0^2 =  {c_{s0}^{\rm eff}}^2$. Therefore, the fact that the critical points coincide with the acoustic horizon further implies that the critical points are the transonic points with effective sound speed given by $ c_{s0}^{\rm eff} = c_{s0}/\sqrt{1+\beta} $. Hence, the apparent mismatch of the critical point and sonic point is resolved if we abandon the static sound speed $ c_s $ and use effective speed of sound $ c_{s0}^{\rm eff} $ as the speed of propagation of acoustic perturbation and define the Mach number as the ratio of the dynamical bulk velocity $ u_0 $ and the effective sound speed $ c_{s0}^{\rm eff} $. In such a case, the critical point and the sonic point (where the Mach number is unity) becomes the same. \\

\noindent The acoustic null ray travelling in the radial direction would be given by $ ds^2|_{ \theta,\phi = \rm constant}  = 0$ \citep{Visser1998}. Thus for acoustic null geodesic, which describes the path of radially travelling phonons, we have 
\begin{equation}\label{null-geodesic}
G_{tt} + 2 G_{rt} \left(\frac{dr}{dt}\right) + G_{rr}\left(\frac{dr}{dt}\right)^2 = 0.
\end{equation}
The metric elements $ G_{\mu\nu} $ are expressed in terms of $ u_0, \lambda_0 $ using equation (\ref{transformation}) as
\begin{equation}\label{key}
\begin{aligned}
& G_{\mu\nu} = \frac{k(r)}{\frac{c_{s0}^2}{1+\beta}}\tilde{G}_{\mu\nu},\\
& \tilde{G}_{tt} = u_0^2-\frac{c_{s0}^2}{1+\beta},\\
& \tilde{G}_{tr} = \tilde{G}_{rt} = -\frac{u_0}{1-u_0^2}\left(1-\frac{c_{s0}^2}{1+\beta}\right)\sqrt{\frac{g_{\phi\phi}}{g_{\phi\phi}-\lambda_0^2 g_{tt}}} ,\\
& \tilde{G}_{rr} =\frac{1}{g_{tt}} \frac{c_{s0}^2}{1+\beta}+\frac{1}{g_{tt}(1-u_0^2)}\left(1-\frac{c_{s0}^2}{1+\beta}\right)\frac{g_{\phi\phi}}{g_{\phi\phi}-\lambda_0^2 g_{tt}}.
\end{aligned}
\end{equation}
The null geodesic is independent of the conformal factor and hence we can use $ \tilde{G}_{\mu\nu} $ instead of $ G_{\mu\nu} $ in equation (\ref{null-geodesic}). $ dr/dt $ obtained from equation (\ref{null-geodesic}) is the coordinate speed of the acoustic phonons as observed from infinity. In the large radial distance, the Schwarzschild metric becomes asymptotically flat. In the non-relativistic Newtonian limit, $ g_{\mu\nu}\to \eta^{\mu\nu} $ and $ u_0\ll 1, c_{s0}\ll 1 $ where $ \eta^{\mu\nu} = {\rm diag}(-1,1,r^2,r^2\sin^2\theta) $ is the flat spacetime metric in the polar coordinate. In this limit we have
\begin{equation}\label{key}
\begin{aligned}
& \tilde{G}_{tt} = u_0^2-\frac{c_{s0}^2}{1+\beta},\\
& \tilde{G}_{tr} = \tilde{G}_{rt} = -u_0,\\
& \tilde{G}_{rr} = 1.
\end{aligned}
\end{equation}
Thus equation (\ref{null-geodesic}) becomes
\begin{equation}\label{key}
(u_0^2-\frac{c_{s0}^2}{1+\beta}) -2u_0 \left(\frac{dr}{dt}\right)+\left(\frac{dr}{dt}\right)^2 = 0,
\end{equation}
which could be rewritten as
\begin{equation}\label{key}
\left|\frac{dr}{dt}-u_0\right| = \frac{c_{s0}}{\sqrt{1+\beta}} = c_{s0}^{\rm eff}.
\end{equation}
The above equation implies that the acoustic perturbation moves with an effective speed $ c_{s0}^{\rm eff} = c_{s0}/\sqrt{1+\beta} $ relative to the moving medium.\\

\noindent For the isothermal  case $ \beta =0 $ for NT and RH and $ \beta = c_{s0}^2 $ for ALP. However, $ c_{s0}^2 \ll 1 $ and therefore, $ 1+ \beta \to 1 $ and hence $ c_{s0}^{\rm eff} = c_{s0} $. Therefore, for isothermal case, the critical point and sonic point coincide in non-relativistic Newtonian accretion flow. For the adiabatic equation of state, $ \beta = (\gamma -1)/2 $ for NT and RH and for ALP $ \beta = (\gamma -1)/2 + c_{s0}^2$. Thus,  $1 +\beta \to (\gamma+1)/2$ as $  c_{s0}^2\ll1 $. Therefore, for adiabatic equation of state, effective sound speed is $ c_{s0}^{\rm eff} = \sqrt{2/(\gamma+1)}c_{s0} $ for all the three disc heights.

\section{Concluding remarks}
\noindent For accretion flow maintained in hydrostatic equilibrium along the vertical direction, the Mach number does not become unity at critical points and hence the critical points and the sonic points become different(location -wise). This happens because, for such a disc model, the flow thickness contains the expression of sound speed. The deviation of Mach number from unity is always observed for polytropic accretion because the sound speed is a position dependent variable for polytropic flow. For isothermal accretion, the sound speed is a position independent constant(because of the temperature invariance). For accretion under the influence of the post-Newtonian black hole potentials, the critical and the sonic points are thus identical for isothermal flow.\\

\noindent The situation is observed to be completely different for complete general relativistic flow. For certain expressions of disc thickness, the Mach number deviates from unity at the critical point even for isothermal accretion. For other disc heights, the critical and the sonic points remain the same for isothermal flow. We try to explain such finding in the following way.\\

\noindent The expression for the flow thickness as obtained by ALP has been derived by setting an energy-momentum conservation equation along the vertical direction as well, in addition to the conservation of the Euler equation along the radial direction (for the equatorial plane). Hence the variation of the sound speed gets intrinsically included in the set of equations (written for the equatorial plane) through the process of vertical averaging, even if one considers the isothermal flow. The sound speed remains position independent constant only along radial direction if such vertical averaging would not be performed. Hence for accretion discs with flow thickness as expressed by ALP, the critical points and the sonic points are formed at different radial distances. For two other expressions of disc heights, the relativistic Euler equation is not constructed or solved along the vertical direction.\\

\noindent The effective dynamical sound speed for different disc models and equation states is given by $ c_{s0}^{\rm eff} = c_{s0}/\sqrt{1+\beta} $. As given in Table \ref{beta}, $ \beta $ depends on the model as well as the equation of state. In particular, it is noticed that for isothermal equation state, $ \beta $ for NT and RH model are zero. For NT and RH, the disc height can be written as $ H = \sqrt{p/\rho}f(r) $ where $ f(r) $ is function of the radial coordinate only. For isothermal equation of state $ p\propto \rho $ and therefore $ H \propto f(r) $. Thus for these two models, the disc height is just a function of $ r $ for the isothermal equation of state. Thus, the height for such cases does not depend on any flow variables such as the velocity or density. Therefore, such prescription of disc height does not make the critical point different from the sonic point.\\

\begin{figure}[h]
  \centering
  \includegraphics{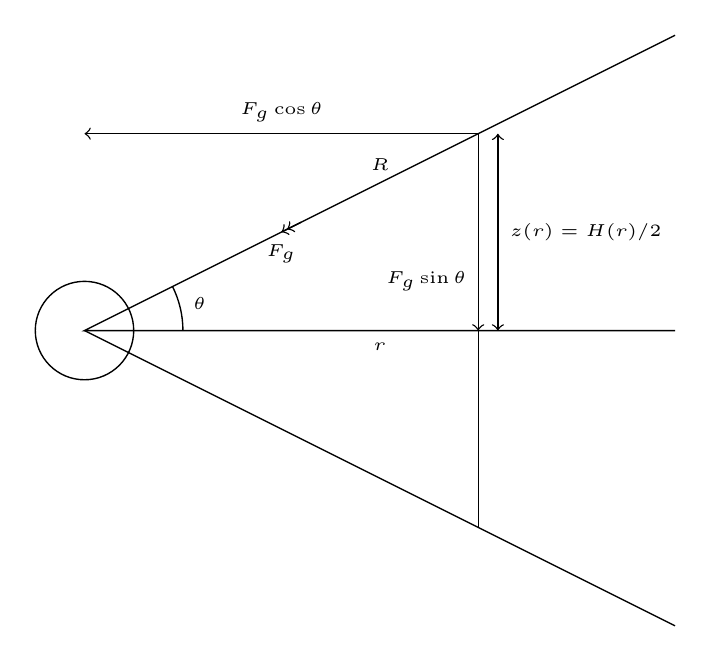}
  \caption{Schematic diagram showing the components of the gravitational force $F_g = d\Phi/dR$ .}
  \label{fig:disc-height}
\end{figure}

\noindent For axially symmetric accretion flow maintained in hydrostatic equilibrium along the vertical direction, the local disc thickness $H(r)$ is a function of the radial sound speed as well. Presence of the sound speed, especially when the sound speed is position dependent, is the prime reason behind the formation of the sonic point at a different location than that of the critical point. The assumption of hydrostatic equilibrium along the vertical direction demands the disc to be geometrically thin. Within the framework of the Newtonian gravity, the thickness can be evaluated using the following procedure. The pressure gradient along the vertical direction is balanced by the component of the gravitational force along that direction. From figure \ref{fig:disc-height}, this gives
\begin{equation}
  \label{eq:garvity-pressure-balance}
  \frac{1}{\rho}\frac{dp}{dz} = \frac{d\Phi}{dR}\times \sin \theta,
\end{equation}
where $\Phi(R)$ is the gravitational potential and $\theta$ is the angle made by the disc at a distance $r$ along the equatorial plane.
Assuming the disc to be thin, i.e., $z(r)\ll r$, where $z(r)$ is the half-thickness of the disc as shown in figure \ref{fig:disc-height}, the above equation becomes
\begin{equation}
  \frac{1}{\rho}\frac{dp}{dz} = \frac{d\Phi}{dr}\times \frac{z(r)}{r},
\end{equation}
The above equation is further approximated as
\begin{equation}
  \frac{1}{\rho}\frac{p}{z(r)} = \frac{d\Phi}{dr}\times \frac{z(r)}{r}.
\end{equation}
For adiabatic equation of state, the sound speed is given by $c_s^2 = \gamma p/\rho$. Thus the disc half-thickness is given by
\begin{equation}
  \label{eq:height}
  z(r) = c_s \sqrt{\frac{r}{\gamma \frac{d\Phi}{dr}}}.
\end{equation}\\

      \noindent While writing $dp/dz$ to be $p/z$, it has been assumed that a differential form can safely be approximated, at least in the present context. Such approximation assumes that the pressure is a (very) slowly varying function of the coordinate associated with the vertical direction.  It is difficult to comment on how accurate such approximation is. To obtain the exact $z$ dependence of the pressure, one needs to formulate and solve the Euler equation along the $z$ direction. Instead of accomplishing such task, usual literature uses the value of the vertically averaged pressure evaluated on the equatorial plane only, thus makes the model effectively one dimensional. Such approximation is probably the major cause behind having a mismatch between the location of the critical and the sonic point. Had it been the case that one would solve a two-dimensional disc structure, the sonic surface would probably coincide with the critical surface. In that case, however, the problem would not be analytically treatable, not even a semi-analytical method would suffice to address the problem, and it would be imperative to take recourse to full numerical solutions.\\

      \noindent Also, it is important to note that a thin disc, where the hydrostatic equilibrium along the vertical direction may be assumed, would incorporate small sound crossing time, and hence, such disc model is more suitable for subsonic flow only. The aforementioned discussion indicates the possible reasons for which the sonic and the critical points are not isomorphic for one-dimensional flow solutions in Newtonian gravity. For general relativistic fluid, the governing equations look more complex and the overall solution method is rather involved. However, the overall underlying logic used to develop the solution remains the same. It is not possible to analytically/semi-analytically construct a two-dimensional disc model where the sonic points would automatically coincide with the critical points. Analytical methods restrict us to use the effective one-dimensional flow structure with vertically averaged values of accretion variables. Within such set of constraints, what best can be done is to redefine the concept of the sound speed through a dynamical approach and to introduce an effective sound speed which makes a sonic point to coincide with a critical point. We have done the same in the present work.\\

\noindent It is, however, important to note that it is difficult to conclude the universality of such phenomena (which kind of disc height will or will not exhibit the non-isomorphism of critical points) since the corresponding expression for flow thicknesses in the works considered here have been derived using a certain set of idealized assumptions. A more realistic flow thickness may be derived by employing the non-LTE radiative transfer \citep{Hubney1998,Davis2006} or by taking recourse to the Grad-Shafranov equations for the MHD flows \citep{Beskin1997,Beskin2005,Beskin2010}.\\

\noindent In the present work, we have obtained the value of the effective sound speed for non-rotating black holes. However, astronomers believe that most of the astrophysical black holes possess non-zero Kerr parameter $ a $ (black hole spin) \citep{Miller2009,Kato2010,Ziolkowski2010,Tchekhovskoy2010,
    Daly2011,Buliga2011,Reynolds2012,McClintock2011,Martínez-Sansigre2011,Dauser2010,Nixon2011,McKinney2012,
    McKinney2013,Brenneman,Rolando2013,Sesana,Fabian2014,
    Healy,Jiang,Nemmen}. Hence in our next work, we would like to perform similar work for
      accretion in the Kerr metric to understand how the black hole spin influences the value of the effective dynamical sound speed.\\

\section{Acknowledgments}
\noindent MAS and SM acknowledge the visit at Sarojini Naidu College Kolkata, and SN acknowledges the visit at HRI (supported by the Cosmology and High Energy Astrophysics Grant). TKD acknowledges the support from the Physics and Applied Mathematics Unit, Indian Statistical Institute, Kolkata, India, in the form of a long-term visiting scientist (one-year sabbatical visitor). The authors would like to thank the anonymous referee for very useful comments and suggestions. 

\section{References}

\bibliographystyle{elsarticle-harv}
\bibliography{reference_arif}

\end{document}